\documentclass[twocolumn,aps,prl,superscriptaddress]{revtex4-2}
\usepackage[utf8]{inputenc}
\usepackage{amsmath}
\usepackage{stix2}
\usepackage{bm}
\usepackage{xcolor}
\usepackage{graphicx}

\begin{document}

\title{
Deep Band Crossings Enhanced Nonlinear Optical Effects
}
\author{Nianlong \surname{Zou}}
\thanks{These authors contributed equally to this work.}
\affiliation{State Key Laboratory of Low Dimensional Quantum Physics and Department of Physics, Tsinghua University, Beijing, 100084, China}

\author{He \surname{Li}}
\thanks{These authors contributed equally to this work.}
\affiliation{State Key Laboratory of Low Dimensional Quantum Physics and Department of Physics, Tsinghua University, Beijing, 100084, China}
\affiliation{Institute for Advanced Study, Tsinghua University, Beijing 100084, China}

\author{Meng \surname{Ye}}
\affiliation{Graduate School of China Academy of Engineering Physics, Beijing, 100193, China}

\author{Haowei \surname{Chen}}
\affiliation{State Key Laboratory of Low Dimensional Quantum Physics and Department of Physics, Tsinghua University, Beijing, 100084, China}

\author{Minghui \surname{Sun}}
\affiliation{State Key Laboratory of Low Dimensional Quantum Physics and Department of Physics, Tsinghua University, Beijing, 100084, China}

\author{Ruiping \surname{Guo}}
\affiliation{State Key Laboratory of Low Dimensional Quantum Physics and Department of Physics, Tsinghua University, Beijing, 100084, China}
\affiliation{Institute for Advanced Study, Tsinghua University, Beijing 100084, China}

\author{Yizhou \surname{Liu}}
\affiliation{School of Physics Science and Engineering, Tongji University, Shanghai 200092, China}

\author{Bing-Lin Gu}
\affiliation{State Key Laboratory of Low Dimensional Quantum Physics and Department of Physics, Tsinghua University, Beijing, 100084, China}
\affiliation{Institute for Advanced Study, Tsinghua University, Beijing 100084, China}

\author{Wenhui \surname{Duan}}
\affiliation{State Key Laboratory of Low Dimensional Quantum Physics and Department of Physics, Tsinghua University, Beijing, 100084, China}
\affiliation{Institute for Advanced Study, Tsinghua University, Beijing 100084, China}
\affiliation{Frontier Science Center for Quantum Information, Beijing, China}

\author{Yong \surname{Xu}}
\email{yongxu@mail.tsinghua.edu.cn}
\affiliation{State Key Laboratory of Low Dimensional Quantum Physics and Department of Physics, Tsinghua University, Beijing, 100084, China}
\affiliation{Frontier Science Center for Quantum Information, Beijing, China}
\affiliation{RIKEN Center for Emergent Matter Science (CEMS), Wako, Saitama 351-0198, Japan}

\author{Chong \surname{Wang}}
\email{chongwang@mail.tsinghua.edu.cn}
\affiliation{State Key Laboratory of Low Dimensional Quantum Physics and Department of Physics, Tsinghua University, Beijing, 100084, China}

\begin{abstract}
Nonlinear optical (NLO) effects in materials with band crossings have attracted significant research interests due to the divergent band geometric quantities around these crossings. Most current research has focused on band crossings between the valence and conduction bands. However, such crossings are absent in insulators, which are more relevant for NLO applications. In this work, we demonstrate that NLO effects can be significantly enhanced by band crossings within the valence or conduction bands, which we designate as ``deep band crossings'' (DBCs). As an example, in two dimensions, we show that shift conductivity can be substantially enhanced or even divergent due to a mirror-protected ``deep Dirac nodal point''. In three dimensions, we propose GeTe as an ideal material where shift conductivity is enhanced by ``deep Dirac nodal lines''. The ubiquity of this enhancement is further confirmed by high-throughput calculations. Other types of DBCs and NLO effects are also discussed. By manipulating band crossings between arbitrary bands, our work offers a simple, practical, and universal way to enhance NLO effects greatly.
\end{abstract}

\maketitle

\textit{Introduction}---Nonlinear optical (NLO) effects are intricately related to the geometric properties of the band structure, such as the Berry connection, the Berry curvature, and the quantum geometric tensor~\cite{bhalla2022resonant,ahn2022riemannian,nagaosa2017concept,morimoto2016topological}. These quantities exhibit a quantitative dependency on the band structure details, which presents challenges in identifying universal patterns in the relationship between NLO effects and band dispersions. However, in materials with band crossings, band geometric quantities are concentrated (and sometimes divergent) around these crossings, making it possible to predict a trend for NLO effects in these materials. For instance, the injection current near the Weyl point is directly related to the Weyl charge~\cite{de2017quantized,yang2017divergent,parker2019diagrammatic}. Divergent shift conductivity (SC) has also been proposed in Dirac and Weyl semimetals~\cite{yang2017divergent,ahn2020low}. In addition, NLO effects of various emergent quasiparticles beyond Weyl and Dirac fermions~\cite{bradlyn2016beyond,yu2022encyclopedia,fang2016topological} have also garnered significant interests~\cite{schilling2017flat,flicker2018chiral,ni2020linear,jiang2023giant,martin2018parity}.

The majority of the aforementioned investigations have primarily focused on band crossings between the valence bands (VBs) and conduction bands (CBs). However, such crossings are absent in insulators, which are more relevant for NLO applications~\cite{Wang2024Aug}. Instead, band crossings within VBs/CBs [designated as ``deep band crossings'' (DBCs)] are abundant in insulators. DBCs have been overlooked in topological studies as they do not affect topological numbers determined by the occupied Bloch state manifold. However, most NLO effects involve processes that engage more than two bands. For these effects, the geometric quantities within the VBs/CBs cannot be overlooked. Therefore, it is natural to anticipate a significant impact on NLO from DBCs.

In this article, we systematically analyze the divergences of band geometric quantities arising from DBCs in second-order NLO responses. By employing the L\"owdin partitioning technique, we establish a comprehensive framework for investigating the divergent Berry connection within VBs/CBs (designated as virtual transitions~\cite{chen2024enhancing,zhang2018photogalvanic,jin2024peculiar}) induced by DBCs in three-band $\mathbf{k} \cdot \mathbf{p}$ models. These divergences can substantially enhance the SC. As a realistic example, we identify GeTe as an ideal material where SC is enhanced by Dirac nodal lines within the CBs. More generally, the ubiquity of this enhancement is confirmed by our high-throughput calculations. Other types of DBCs and NLO effects are also discussed. By engineering band crossings between arbitrary bands, such as through pressure-induced structural transition, our work offers a simple, practical, and universal way to enhance NLO effects.

\begin{figure}
    \includegraphics[width=\linewidth]{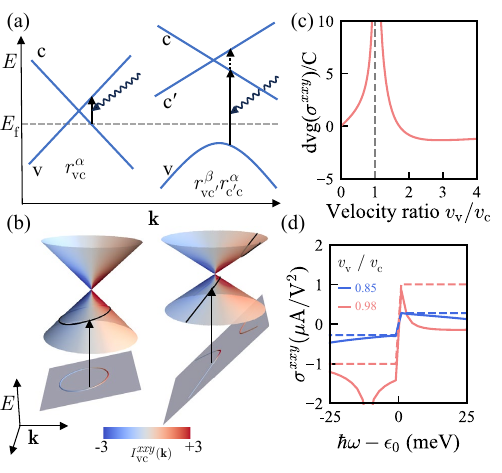}\textbf{}
    \caption{
    (a) A schematic diagram illustrating the divergent real transition ($r^\alpha_{\mathrm{vc}}$) produced by band crossings between the conduction and valence bands, and the divergent virtual transition ($r^\alpha_{\mathrm{c'c}}$) generated by DBCs. (b) A schematic representation of optical type-I and type-II Dirac points, with the resonant transition subspace indicated by black lines. The color indicates the magnitude of \(I_\mathrm{vc}^{xxy}\) at the corresponding $\mathbf{k}$ point. (c) dvg($\sigma^{xxy}$) [cf.~Eq.~\eqref{integral}] of the mirror nodal point as a function of velocity ratio $v_{\mathrm{v}}/v_{\mathrm{c}}$, showing a divergent behavior as the ratio approaches 1. (d) $\sigma^{xxy}$ (solid lines) and dvg($\sigma^{xxy}$) for the tight-binding models of nodal points in the optical type-I regime. $\sigma^{xxy}$ is obtained by numerical integration and dvg($\sigma^{xxy}$) is an analytic approximation to $\sigma^{xxy}$.
    }
    \label{fig1}
\end{figure}

\textit{Divergence in virtual transitions}---In the vicinity of band crossing points (or lines or surfaces), geometric quantities associated with degenerate bands often exhibit divergent behavior, such as the Berry curvature near a Weyl point. As a Weyl point is approached, the Berry curvature $\mathbf{\Omega}$ diverges as $1/|\mathbf{k}-\mathbf{k}_0|^2$, where $\mathbf{k}_0$ is the position of the Weyl point in the Brillouin zone (BZ). In a simplified two-band approximation (bands labeled as 1, 2), the divergence of the Berry curvature can be attributed to the divergence of the interband Berry connection $\mathbf{r}_{12}$: $\Omega^\gamma_1 = i \varepsilon^{\alpha \beta \gamma} r^{\alpha}_{12} r^{\beta}_{21}$, where $\varepsilon^{\alpha \beta \gamma}$ represents the Levi-Civita symbol and \{$\alpha$, $\beta$, $\gamma$\} are Cartesian indices. Essentially, most geometric quantities are composed of the Berry connection and its derivatives. Hence, the divergent interband Berry connection serves as the primary source of divergent geometric quantities. The relationship between the divergence of the interband Berry connection and the corresponding band energy degeneracy can be revealed by a well-known equation~\cite{sipe2000second}:
\begin{equation}\label{rnm}
r_{nm}^\alpha= -i \frac{v_{nm}^\alpha}{\omega_{nm}} =-i \frac{\langle n |\partial_{k_\alpha} H(\mathbf{k}) | m \rangle}{\omega_{nm}}.
\end{equation}
where $H(\mathbf{k})$ is the Bloch Hamiltonian and \{$n$, $m$\} denote band indices. Here, $\omega_{nm} = (E_n - E_m) / \hbar$, with $E_n$ and $E_m$ being corresponding band energies. Note that the velocity matrix element $v_{nm}^\alpha$ cannot diverge due to the normalization of the eigenstates and the analytic nature of the Hamiltonian. Therefore, the divergence of $r_{nm}^\alpha$ can only occur when the two corresponding bands, $n$ and $m$, become degenerate at $\mathbf{k}_0$. Consequently, for optical effects only involving the Berry connection between valence and conduction bands (designated as real transitions), such divergence will occur exclusively in semimetals where a band crossing is present near the Fermi surface [see Fig.~\ref{fig1}(a)]. Furthermore, a straightforward power counting analysis reveals that whenever $r_{nm}^\alpha$ diverges, it scales as $1/|\mathbf{k}-\mathbf{k}_0|$, given that $r_{nm}^\alpha$ has the dimension of length~\cite{ahn2020low}.

In the realm of optical responses involving more than two bands, the divergence of the virtual transitions within CBs or VBs can become significant. To elucidate the fundamental principles of DBCs, we begin with a three-band Hamiltonian. This Hamiltonian comprises two nearly degenerate CBs (denoted by $|\mathrm{c}\rangle$ and $|\mathrm{c'}\rangle$) and one VB (denoted by $|\mathrm{v}\rangle$) [see Fig.~\ref{fig1}(a)]:
\begin{equation}\label{kpblock}
H(\mathbf{k}) = \left(
\begin{array}{ccc}
\mathbf{h}(\mathbf{k})\cdot \bm{\sigma} & \mathbf{t}(\mathbf{k}) \\
\mathbf{t}^\dagger (\mathbf{k})   & \epsilon(\mathbf{k})
\end{array}
\right),
\end{equation}
where $\mathbf{h}(\mathbf{k})\cdot \bm{\sigma}$ is a two-by-two matrix describing the two CBs, $\mathbf{t}(\mathbf{k})$ is a vector representing the coupling between the CBs and the VB, and $\epsilon(\mathbf{k})$ represents the VB dispersion without conduction-valence band coupling. In a certain region of the BZ, $\mathbf{h}=\mathbf{t}=\bm{0}$, leading to the crossing of the two CBs. The dimension of this crossing is currently not specified. Around this crossing, the Berry connection between $|\mathrm{c}\rangle$ and $|\mathrm{c'}\rangle$ may exhibit divergent behavior. Such divergences are commonly encountered in NLO effects and are typically non-physical, resulting in an ultimate cancellation. However, in a specific class of NLO effects where direct current is generated, these divergences do not cancel. We illustrate this phenomenon using the shift current as an example.

\textit{Enhancement in the shift current}--The shift current is a second-order optical response that characterizes the generation of a direct current by incident light~\cite{sipe2000second,wang2017first,wang2019first}. The central object in the expression of the SC is a band geometric quantity $I^{\alpha \beta \gamma}_{mn} = \mathrm{Im}(r_{mn}^{\beta} D^{\alpha} [ r_{nm}^{\gamma} ] + r_{mn}^{\gamma} D^{\alpha} [ r_{nm}^{\beta} ])$. Here, \(D^{\alpha} [ r_{nm}^{\gamma} ]\) represents the $U(1)$ covariant derivative defined as \(D^{\alpha} [ r_{nm}^{\gamma} ] = \partial_{k_\alpha} r^{\gamma}_{nm} - i(A^{\alpha}_{n} - A^{\alpha}_{m})r^{\gamma}_{nm}\), where \(A^{\alpha}_{n}\) and \(A^{\alpha}_{m}\) are the intraband Berry connections of bands \(n\) and \(m\), respectively. Near DBCs ($\omega_\mathrm{cc'}\approx 0$), the covariant derivative \(D^{\alpha} [ r_\mathrm{cv}^{\gamma} ]\) may diverge due to possibly diverging virtual transitions. By employing a sum rule~\cite{cook2017design,aversa1995nonlinear}, the divergent part of the covariant derivative is
\begin{equation}\label{dvg1}
\text{dvg}( D^{\alpha} [ r_{\mathrm{cv}}^{\gamma} ])=i r_{\mathrm{cc'}}^{\alpha}r_{\mathrm{c'v}}^{\gamma}
\end{equation}
for the three-band model defined in Eq.~\eqref{kpblock}.

As depicted in Fig.~1(a), $\text{dvg}( D^{\alpha} [ r_{\mathrm{cv}}^{\gamma} ])$ can be interpreted as a combination of a real optical transition from $\mathrm{c}'$ to $\mathrm{v}$ and a virtual transition from $\mathrm{c}$ to $\mathrm{c}'$. Since the divergence primarily originates from the virtual transition $r^\alpha_{\mathrm{cc'}}$, which only contains information about the nearly degenerate CBs, it should be possible to analyze the divergent behavior within this subspace. Using the L\"owdin partitioning technique, the divergent part of $I^{\alpha \beta \gamma}_{\mathrm{vc}}$ is given by (see the derivation in Supplemental Material~\cite{SM}):
\begin{equation}\label{dvg}
\mathrm{dvg}(I^{\alpha \beta \gamma}_{\mathrm{vc}}) = \frac{1}{2\epsilon^2(\mathbf{k}_0)}\mathrm{Re}[\partial_{k_\beta} \mathbf{t}^\dagger (\mathbf{k}) \cdot (\mathbf{\tilde{h}}\times \partial_{k_\alpha} \mathbf{\tilde{h}}
\cdot \hat{\bm{\sigma}}) \cdot \partial_{k_\gamma} \mathbf{t}(\mathbf{k})],
\end{equation}
where $\mathbf{\tilde{h}}=\mathbf{h}/|\mathbf{h}|$.

In Eq.~\eqref{dvg}, $\partial_{k_\beta} \mathbf{t}$ and $\partial_{k_\gamma} \mathbf{t}$ result from real transitions between VB and CBs and do not diverge. The potentially divergent term is the cross product $\mathbf{\tilde{h}}\times \partial_{k_\alpha} \mathbf{\tilde{h}}$, which possesses a clear geometric interpretation. Since that $\mathbf{\tilde{h}}(\mathbf{k})$ defines a mapping from the $\mathbf{k}$-space to the Bloch sphere, $\mathbf{\tilde{h}}(\mathbf{k})$ can be regarded as a trajectory on the Bloch sphere. Consequently, $\mathbf{\tilde{h}}\times \partial_{k_\alpha} \mathbf{\tilde{h}}$ represents the ``angular velocity'' of this trajectory. In the near-degenerate region, the angle of $\mathbf{\tilde{h}}$ undergoes a significant change within a small interval of the $\mathbf{k}$-space, potentially leading to divergence.

\begin{figure*}
    \includegraphics[width=\linewidth]{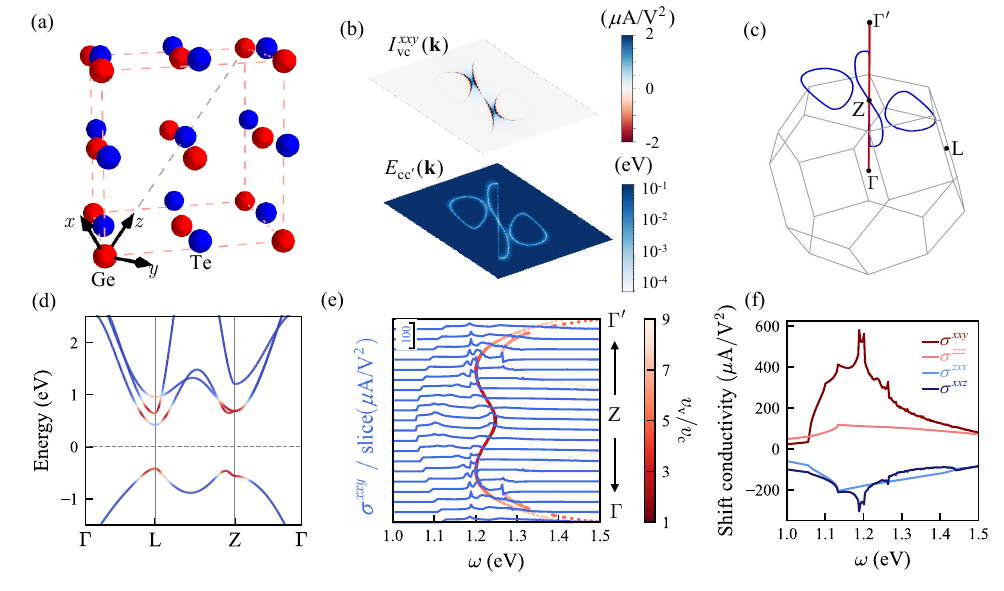}\textbf{}

    \caption{(a) The crystallographic structure of GeTe, with the z-axis (grey dashed line) oriented along the diagonal of the unit cell. (b) A comparison between the energy difference $E_{\mathrm{cc'}}$ (bottom) and the integrand $I^{xxy}_{\mathrm{vc}}$ (top). (c) Depiction of the BZ and nodal lines in GeTe, with the red (blue) line indicating the nodal line protected by $C_{3v}$ ($M_x$). (d) The band structure of GeTe in the absence of SOC, highlighting in red the bands contributing to the SC peak at 1.18~eV. (e) $\sigma^{xxy}$ as defined in Eq.~\eqref{sc} but the integral is restricted in a small interval in $k_z$. At each $k_z$, the nodal line is fitted to the two-dimensional nodal point model, with the energy parameter $\epsilon_0$ from this model depicted by the red curve. The corresponding velocity ratio $v_\mathrm{v} / v_\mathrm{c}$ is indicated by the intensity of this curve. (f) The SC of GeTe calculated along four symmetry-permitted directions.
    }
    \label{fig2}
\end{figure*}

Having identified the $\mathbf{k}$-space divergence of $I^{\alpha \beta \gamma}_{mn}$, we proceed to investigate its effect on the transient SC~\cite{belinicher1982kinetic,sturman2020ballistic,alexandradinata2023anomalous}, defined as follows:
\begin{equation}\label{sc}
  \sigma^{\alpha \beta \gamma} = \frac{\pi e^3}{\hbar^2} \int \frac{\mathrm{d} \mathbf{k}}{(2 \pi)^3} \sum_{n, m} f_{nm} I_{mn}^{\alpha \beta \gamma} \delta (\omega_{nm} - \omega),
\end{equation}
where \(\omega\) denotes the frequency of the incident light, and \(f_{nm} = f_n - f_m\) with \(f_n\) and \(f_m\) being the Fermi distribution functions. To provide a concrete example, we utilize a specific two-dimensional \(\mathbf{k} \cdot \mathbf{p}\) model that characterizes a nodal point protected by mirror symmetry in the $x$ direction ($M_x$): \(\mathbf{h}(\mathbf{k}) = (a k_x, b k_x, v_{\mathrm{c}} k_y)\), \(\mathbf{t}(\mathbf{k}) = (d_y k_y, d_x k_x)\), and \(\epsilon(\mathbf{k}) = \epsilon_0 + v_{\mathrm{v}} k_y\) [cf.~Eq.~(\ref{kpblock})]. In this model, the two CBs have opposite $M_x$ eigenvalues, forming a nodal point at the origin. \(\epsilon(\mathbf{k})\) represents a trivial band that tilts along the \(k_y\) axis, and \(\mathbf{t}(\mathbf{k})\) is the symmetry-allowed coupling between the CBs and VB. After substituting above expressions into Eq.~\eqref{dvg}, it becomes apparent that only $I_{\mathrm{vc}}^{xxy}$ and $I_{\mathrm{vc}}^{yxy}$ exhibit divergent behavior, expressed as follows: dvg$(I^{xxy}_{\mathrm{vc}})=-v_{\mathrm{c}} b d_x d_y k_y/(2\epsilon_0^2 |\mathbf{h}|^2)$ and dvg$(I^{yxy}_{\mathrm{vc}})=v_{\mathrm{c}} b d_x d_y k_x/(2\epsilon_0^2 |\mathbf{h}|^2)$. These divergences are validated numerically in Supplemental Fig. S1~\cite{SM}.

We denote the contribution of $\mathrm{dvg}(I_{\mathrm{vc}}^{\alpha \beta \gamma})$ to $\sigma^{\alpha \beta \gamma}$ as $\mathrm{dvg}(\sigma^{\alpha \beta \gamma})$. Due to the $\mathbf{k}$-space integration in Eq.~(\ref{sc}), $\mathrm{dvg}(\sigma^{\alpha \beta \gamma})$ does not necessarily diverge. However, it represents a significant contribution to the SC due to the divergence of $\mathrm{dvg}(I_{\mathrm{vc}}^{\alpha \beta \gamma})$. In certain cases (specified below), $\mathrm{dvg}(\sigma^{\alpha \beta \gamma})$ does diverge, and it represents the sole divergent part of $\sigma^{\alpha \beta \gamma}$ in our three-band model.

Since $\sigma^{yxy}$ is forbidden by $M_x$ symmetry~\cite{gallego2019automatic} , the divergence in dvg$(I^{yxy}_{\mathrm{vc}})$ cancels out after integration in Eq.~(\ref{sc}). Therefore, we focus on the $\sigma^{xxy}$ component. The Hamiltonian of the nodal point model is linear and is simple enough such that $\mathrm{dvg}(\sigma^{xxy})$ can be obtained analytically. The frequency dependence of $\mathrm{dvg}(\sigma^{xxy})$ can be expressed as a sign function after the scaling (or change of variable) $\tilde{k}_y = {v_{\mathrm{c}}k_y}/{(\hbar\omega-\epsilon_0)}$ and $\tilde{k}_x = {\sqrt{a^2+b^2}k_x}/{(\hbar\omega-\epsilon_0)}$:
\begin{equation}\label{integral}
\mathrm{dvg}(\sigma^{xxy}) = C \cdot \mathrm{sign}(\hbar\omega-\epsilon_0) \int \frac{\tilde{k}_y}{|\mathbf{\tilde{k}}|^2}[\delta(\varsigma^+)+\delta(\varsigma^-)]\mathrm{d}\mathbf{\tilde{k}},
\end{equation}
where $C = {e^2 b d_x d_y}/{(8 \pi \hbar v_{\mathrm{c}}\epsilon_0^2\sqrt{a^2+b^2})}$, $\varsigma^{\pm}=\pm|\mathbf{\tilde{k}}|-(v_{\mathrm{v}}/v_{\mathrm{c}}) \tilde{k}_y-1$. The integral then depends solely on the velocity ratio $v_{\mathrm{v}}/v_{\mathrm{c}}$. Analogous to the classification of type-I and type-II Dirac points, two distinct scenarios emerge [as illustrated in Fig.~1(b)]: if $v_{\mathrm{v}} < v_{\mathrm{c}}$, the optical transition surface is elliptical, referred to as ``optical type-I''; if $v_{\mathrm{v}} > v_{\mathrm{c}}$, the optical transition surface becomes hyperbolic, referred to as ``optical type-II''.

Equation~(\ref{integral}) demonstrates a significant enhancement of $\sigma^{xxy}$ due to the Dirac nodal point. Especially, when \(v_{\mathrm{v}} = v_{\mathrm{c}}\), $\mathrm{dvg}(\sigma^{xxy})$ diverges, as illustrated in Fig.~1(c). To investigate the divergence of dvg(\(\sigma^{xxy}\)) in a more realistic context, we construct a 3-band tight-binding model featuring a Dirac nodal point protected by $M_x$ symmetry~\cite{SM}). In Fig.~1(d), we compare the numerical evaluation of $\sigma^{xxy}$ for the tight-binding model with dvg(\(\sigma^{xxy}\)) calculated from Eq.~(\ref{integral}). A good agreement between the two is generally observed. However, the frequency dependence of dvg(\(\sigma^{xxy}\)) for the tight-binding model near \(v_{\mathrm{v}} = v_{\mathrm{c}}\) is no longer a sign function. The underlying reason is that the integral is unbounded in Eq.~(\ref{integral}), while the integral of Eq.~(\ref{sc}) for the tight-binding model is restricted to the BZ. A subsequent analysis indicates that $\mathrm{dvg}(\sigma^{xxy})$ should behave as $1/(\hbar\omega - \epsilon_0)$ near \(v_{\mathrm{v}} \approx v_{\mathrm{c}}\) for the tight-binding model~\cite{SM}. In this case, the divergence of $\sigma^{xxy}$ is retained at $\hbar\omega = \epsilon_0$.

In two-dimensional (2D) materials, finding a Dirac nodal point that satisfies \(v_{\mathrm{v}} = v_{\mathrm{c}}\) usually requires fine-tuning an additional adjustable parameter. In three-dimensional (3D) materials, the crystal momentum \(k_z\) naturally serves as this tunable parameter. Especially, In 3D Dirac nodal line materials~\cite{lau2019topological,fang2016topological}, the nodal line can be viewed as a continuous collection of 2D Dirac points, where the parameters of these Dirac points vary with $k_z$. Here, we report GeTe as a model material to explore the DBC-enhanced SC arising from Dirac nodal lines (Fig.~\ref{fig2}).

\textit{Giant shift conductivity in GeTe}---As illustrated in Fig.~\ref{fig2}(a), the crystal structure of the ferroelectric phase of GeTe resembles the rocksalt structure but is distorted along the $z$-axis. An additional relative shift between the Ge and Te sublattices breaks the inversion symmetry, placing GeTe in the space group \(C_{3v}\) with the \(C_3\) axis aligned along the $z$ direction. In $xxy$ and $xxz$ directions, GeTe exhibits a giant SC peak that reaches 580 $\mathrm{\mu A / V^2}$ at 1.18~eV~[Fig.~\ref{fig2}(f)]. This peak is primarily contributed by optical transitions involving two CBs and a VB, highlighted in red in Fig.~2(d), making GeTe an ideal 3-band model material. The two CBs form a complex nodal line structure in the BZ [Fig.~2(c)], where two types of nodal lines can be identified: the red line along the $\Gamma$-$Z$ direction is protected by \(C_{3v}\) symmetry and the blue lines on the mirror plane are accidental band crossings formed by bands with opposite mirror eigenvalues. The different symmetries lead to distinct divergent behaviors. As shown in Fig.~2(b), near the mirror nodal line, the expected divergence is observed, with \(I^{xxy}\) reaching positive or negative infinity as the nodal line is approached. However, due to excessive symmetries, the \(C_{3v}\) nodal line does not exhibit divergence, which will be discussed in Supplemental Material~\cite{SM}.

As mentioned above, a 3D Dirac nodal line can be viewed as a continuous series of 2D Dirac points, parameterized by $k_z$. For each $k_z$, we extract the parameters $\epsilon_0$, $v_{\mathrm{c}}$ and $v_{\mathrm{v}}$. In addition, we integrate $I^{\alpha \beta \gamma}$ over a small slice of the BZ, determined by the interval $[k_z, k_z + \Delta k_z]$. The integration results, together with the velocity ratio $v_{\mathrm{v}} / v_{\mathrm{c}}$, are presented in Fig.~\ref{fig2}(e). Consistent with the 2D model, the main contribution of SC is concentrated in the region where \(v_{\mathrm{v}} / v_{\mathrm{c}}\) approaches 1. However, an additional factor unique to 3D materials is observed. In the interval where \(\epsilon_0\) shows minimal variation along the nodal line, the enhancement effects can accumulate, which can naturally occur at the ``band edge'' of the nodal line. Finally, by summing the SC over different slices, the $xxy$ component of SC is presented in Fig.~2(f) along with other independent components. The peak values of all these components are significantly larger than the peak value of 33 \(\mathrm{\mu A / V^2}\) observed in the traditional semiconductor BaTiO\(_3\)\cite{young2012first}. The enhanced response of the \(zzz\) and \(zxx\) components has been previously attributed to factors such as high covalency and a narrow band gap~\cite{gong2018phonon,tiwari2022enhanced}. However, their peak values are markedly lower than those of the \(xxy\) and \(xxz\) components, which are further amplified by the presence of ``deep Dirac nodal lines.'' This distinctive anisotropy in the SC provides further evidence of the critical impact of DBCs on SC. For simplicity, the effect of spin-orbit coupling (SOC) is neglected in the above calculations; results including SOC effects remain qualitatively unchanged and can be found in the Supplemental Material~\cite{SM}.

\begin{figure}
    \includegraphics[width=\linewidth]{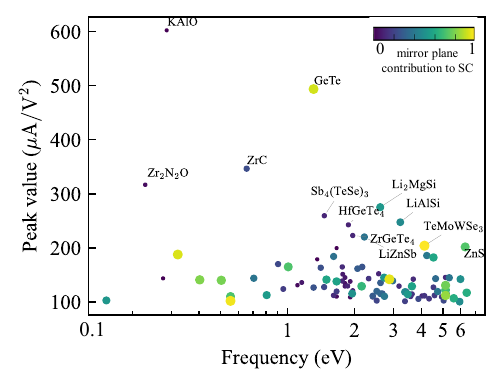}\textbf{}
    \caption{The statistical graph displays the SC peak values and corresponding light frequencies for various materials. For each peak, the color and size of the data points are used together to indicate the ratio of the integral [see Eq.~(\ref{sc})] near the mirror plane to the total integral, thereby reflecting the contribution of possible mirror nodal lines to the SC peak. The calculations include the SOC effect.
    }
    \label{fig3}
\end{figure}

\textit{High-throughput calculations}---To investigate whether SC enhanced by mirror nodal lines is ubiquitous, we performed high-throughput calculations of SC for 2032 mirror-symmetric materials from the Materials Project database~\cite{jain2013commentary}. For materials with SC peak exceeding 100 \(\mathrm{\mu A / V^2}\), Fig.~\ref{fig3} presents their peak values, the corresponding frequencies, and the contribution to the peak from the region near the mirror plane. Notably, for 24\% of materials in Fig.~\ref{fig3}, the region near the mirror plane contributes over 50\% of the SC peak, indicating the widespread occurrence of SC enhancement due to Dirac band crossings (DBCs).

\textit{Discussion}---Thus far, we have examined Dirac points in 2D and Dirac nodal lines in 3D systems, both protected by mirror symmetry. For other types of band crossings, the impact of DBCs on NLO effects is drastically different. Band crossings can be classified by their codimensions $L$, defined as \(L = N - N_{\text{deg}}\), where \(N\) is the system's dimension and \(N_{\text{deg}}\) is the dimension of the degenerate band crossing. For \(L = 0\), the band degeneracy extends across the entire BZ. In this case, the expression of NLO effects must be modified to ensure invariance under $U(2)$ transformations within the band degeneracy space~\cite{chen2022basic}. In this reformulated expression, the virtual transitions within degenerate bands are effectively absent, and thus no divergence of NLO effects will appear. For \(L = 1\) DBCs, the divergent behavior of band geometric quantities can be reduced to the behavior observed in DBCs with other codimensions~\cite{SM}. The most important case is \(L = 2\). Besides mirror-protected nodal lines, we have investigated the divergent behavior of all other nodal lines, including those protected by \(C_{3v}\), \(C_{4v}\), and \(C_{6v}\) symmetries. Due to the excessively high symmetry, the divergent integrands $\text{dvg}(I^{\alpha\beta\gamma}_\mathrm{vc})$ for all these nodal lines vanish. For band crossings with \(L = 3\), such as a Weyl point at $\mathbf{k}_0$, as $\mathbf{k}$ approaches $\mathbf{k}_0$, the divergent virtual transition proportional to \(1/|\mathbf{k}-\mathbf{k}_0|\) will be overwhelmed by the diminishing integral surface element, which is proportional to \(|\mathbf{k}-\mathbf{k}_0|^2\). As a result, the SC will not receive a significant contribution from \(L = 3\) DBCs.

We have examined various other second-order NLO phenomena in the Supplemental Material~\cite{SM}. Magnetic shift current also exhibits terms that diverge in the presence of DBCs. However, for the second harmonic generation, the divergence induced by virtual transitions within the interband components is canceled by those in the intraband components. The detailed analysis for these NLO effects is left for future works.

\begin{acknowledgments}
This work was supported by the Basic Science Center Project of NSFC (Grant No. 52388201), the National Natural Science Foundation of China (Grant No. 12334003, no.12421004, and no.12361141826), the National Science Fund for Distinguished Young Scholars (Grant No. 12025405), the National Key Basic Research and Development Program of China (Grant No. 2023YFA1406400), the Beijing Advanced Innovation Center for Future Chip (ICFC), the Beijing Advanced Innovation Center for Materials Genome Engineering, and the NSAF center project of NSFC No. U2330401. The work was carried out at the National Supercomputer Center in Tianjin using the Tianhe new generation supercomputer.

\end{acknowledgments}

\bibliography{reference}

\end{document}